\documentclass[pra,showpacs,preprintnumbers,twocolumn,superscriptaddress,
amsmath,amssymb]{revtex4}
\usepackage{graphicx}
\usepackage{dcolumn}
\usepackage{bm}
\usepackage[dvips]{color}
\newcommand{\bfk}{{\bf k}}
\newcommand{\ek}{\epsilon_\bfk}
\newcommand{\ef}{E_F}
\newcommand{\cks}{c_{\bfk,\sigma}}
\newcommand{\aks}{a_{\bfk,\sigma}}
\newcommand{\au}{a_{1}}
\newcommand{\ad}{a_{2}}
\newcommand{\cku}{c_{\bfk,\uparrow}}

\newcommand{\cmkd}{c_{-\bfk,\downarrow}}

\newcommand{\bz}{b_0}

\newcommand{\jm}{{\cal J}_-}
\newcommand{\jp}{{\cal J}_+}
\newcommand{\jx}{{\cal J}_x}
\newcommand{\jy}{{\cal J}_y}
\newcommand{\jz}{{\cal J}_z}

\begin{document}
\title{Many-body effects on adiabatic passage through Feshbach
resonances}

\author{I. Tikhonenkov} 
\affiliation{Department of Chemistry, Ben-Gurion University of
the Negev, P.O.B. 653, Beer-Sheva 84105, Israel}
\author{E. Pazy} 
\affiliation{Department of Chemistry, Ben-Gurion University of
the Negev, P.O.B. 653, Beer-Sheva 84105, Israel}
\author{Y. B. Band} 
\affiliation{Department of Chemistry, Ben-Gurion University of
the Negev, P.O.B. 653, Beer-Sheva 84105, Israel}
\author{M. Fleischhauer}
\affiliation{Fachbereich Physik, Technische Universit\"at
Kaiserslautern, D67663, Kaiserslautern, Germany}
\author{A. Vardi}
\affiliation{Department of Chemistry, Ben-Gurion University of
the Negev, P.O.B. 653, Beer-Sheva 84105, Israel}

\date{\today}

\begin{abstract} 
We theoretically study the dynamics of an adiabatic sweep through a
Feshbach resonance, thereby converting a degenerate quantum gas of
fermionic atoms into a degenerate quantum gas of bosonic dimers.  Our
analysis relies on a zero-temperature mean-field theory which
accurately accounts for initial molecular quantum fluctuations,
triggering the association process.  The structure of the resulting
semiclassical phase-space is investigated, highlighting the dynamical
instability of the system towards association, for sufficiently small
detuning from resonance.  It is shown that this instability
significantly modifies the finite-rate efficiency of the sweep,
transforming the single-pair exponential LZ behavior of the remnent
fraction of atoms $\Gamma$ on sweep rate $\alpha$, into a a power law
dependence as the number of atoms increases.  The obtained
nonadiabaticity is determined from the interplay of characteristic
timescales for the motion of adiabatic eigenstates and for fast
periodic motion around them.  Critical slowing-down of these
precessions near the instability, leads to the power-law dependence.
A Linear power-law $\Gamma \propto \alpha$, is obtained when the
initial molecular fraction is smaller than the $1/N$ quantum
fluctuations, and a cubic-root power-law $\Gamma \propto \alpha^{1/3}$
is attained when it is larger.  Our mean-field analysis is confirmed
by exact calculations, using Fock-space expansions.  Finally, we fit
experimental low temperature Feshbach sweep data with a power-law
dependence.  While the agreement with the experimental data is well
within experimental error bars, similar accuracy can be obtained with
an exponential fit, making additional data highly desirable.
\end{abstract}

\maketitle

\section{\label{Sec:intro}Introduction}

Many of the most exciting experimental achievements in ultra-cold
atomic physics in recent years have used Feshbach resonances
\cite{Stwalley76,Tiesinga93, Inouye98, Mies00, Timmermans99}.  Not
only are they a tool for altering the strength and sign of the
interaction energy of atoms \cite{Tiesinga93,Inouye98}, they also
provide a means for converting atom pairs into molecules, and vice
versa \cite{Timmermans99,Regal03, Strecker03, Cubizolles03,Hodby,
Greiner03, Jochim03, Zwierlein03,Claussen02,Herbig03,Durr04}.  A
magnetic Feshbach resonance involves the collisional coupling of free
atom pairs (the asymptotic limit at large internuclear separation, of
the incident open channel molecular potential) in the presence of a
magnetic field, to a bound diatomic molecule state (the closed
channel) on another electronic molecular potential surface.  The
difference in the magnetic moments of the atoms correlating
asymptotically at large internuclear distance to the two potential
energy surfaces, allows the Feshbach resonance to be tuned by changing
the magnetic field strength \cite{Inouye98}.  Sweeping the magnetic
field as a function of time, so that the bound state on the closed
channel passes through threshold for the incident open channel from
above, can produce bound molecules.  This technique has proved to be
extremely effective in converting degenerate atomic gases of fermions
\cite{Regal03, Strecker03,Cubizolles03, Hodby, Greiner03,
Jochim03,Zwierlein03} and bosons \cite{Claussen02, Herbig03,Durr04}
into bosonic dimer molecules.  Fermions are better candidates for
Feshbach sweep experiments due to the relatively long lifetimes of the
resulting bosonic molecules, originating from Pauli blocking of
atom-molecule and molecule-molecule collisions when the constituent
atoms are fermions \cite{Petrov}.  

Here we consider the molecular production efficiency of adiabatic
Feshbach sweep experiments in Fermi degenerate gases.  We determinine
the functional dependence of the remnent atomic fraction $\Gamma$ on
the Feshbach sweep rate $\alpha$, following the treatment in a
previous Letter \cite{Pazy05}, extending the calculations, and
presenting a more detailed account of the theoretical methodology.

The Fermi energy is the smallest energy scale in the system in the
fermionic Feshbach sweep experiments of Refs.~\cite{Regal03,
Strecker03, Cubizolles03, Hodby,Greiner03,Jochim03,Zwierlein03}.
Hence, we treat the fermions theoretically as occupying the lowest
possible many-body state consistent with symmetry considerations
arising from the method of preparation.  Consequently we assume
that the quantum states are filled up to the Fermi energy in a fashion
consistent with the symmetry properties of the gas.  In this sense,
the gas can be thought of as a zero temperature gas.

\begin{figure}
\includegraphics[scale=0.45,angle=0]{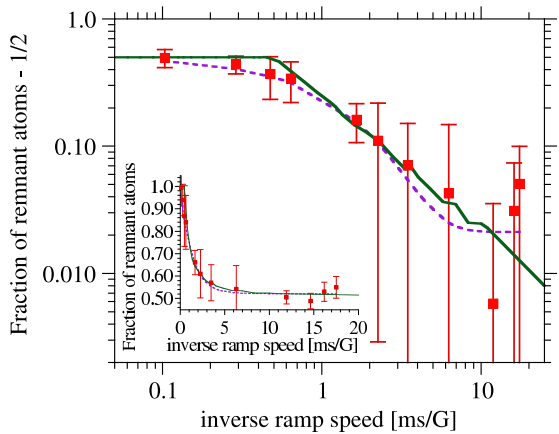}
\caption{(Color online) Fraction of remnant atoms, $\Gamma$, versus inverse 
ramp speed $1/\dot{B}$ across the 543 G resonance of $^6$Li.  The
experimental data (red squares) of \cite{Strecker03}, which saturates
at a remnant fraction of $1/2$ \cite{Pazy04}, and the mean-field
calculations (solid green curve) obey a linear dependence on sweep
rate beyond 0.5 ms/G. ${g^2\over\alpha N}$ is multiplied by 0.5 ms/G
to scale the abscissa for the calculated results.  Also shown as a
dashed blue line in the inset is the best exponential fit to the
data, $\Gamma = 0.479 \exp(-\alpha/1.3) + 0.521$.} \label{fig1}
\end{figure}

The Landau-Zener (LZ) model \cite{Landau} is the paradigm for
explaining how transitions occur in the collision of a single pair of
atoms in a Feshbach sweep experiment.  Theoretical analysis of
molecular production efficiency in Feshbach sweep experiments in a gas
phase have been based on Landau-Zener theory \cite{Mies00}.
Exponential fits have also been carried out for experimental molecular
efficiency data.  Fig.~\ref{fig1} shows data from 
experiments on a quantm degenerate gas of $^6$Li atoms 
\cite{Strecker03}, plotting the remaining fraction of atoms 
(red squares) as a function of the inverse magnetic sweep rate.
The inset of Fig.~\ref{fig1} includes an exponential fit 
(blue dashed curve), $\Gamma = 0.479\exp(-\alpha/1.3) + 0.521$,
taken from Ref.~\cite{Strecker03}. While the exponential
curve lies well within all experimental error bars, the data 
fits a linear power-law dependence (green curve) to
the same level of accuracy.  In what follows we provide the
theoretical detail required to obtain the linear power-law fit in
Fig.~\ref{fig1}.  We show that due to the nonlinearity of the reduced
single-particle (i.e. mean-field) description of the many-atom system,
instabilities are made possible.  These instabilities result in the
failure of the standard LZ theory when the number of atoms is large.
We also predict two different power-law behaviors, $\Gamma \propto
\alpha$ and $\Gamma \propto \alpha^{1/3}$, depending on the initial
state of the system prior to the Feshbach sweep.

The paper is organized as follows.  In Section \ref{Sec:Model} we
introduce the Feshbach model-Hamiltonian and the
main approximations used.  In Sec.~\ref{Sec:phase space} we describe
the classical phase space obtained for the above model.  Section
\ref{sec:sweep} describes how the molecular production efficiency
depends on the stability of the fixed points for the equations of
motion.  In Sec.~\ref{sec:quantum_fluctuations} we describe the role
of quantum fluctuations which lead to the linear dependence of the
molecular production efficiency on the sweep rate.  The analysis of
sections \ref{sec:sweep} and \ref{sec:quantum_fluctuations} is
verified by the exact numerical calculations presented in section
\ref{sec:numerical}.  Section \ref{sec:summary} contains summary and
conclusions.

\section{The Zero Temperature Model System} \label{Sec:Model}

Experiments on molecule production in slowly swept, broad Feshbach
resonance systems \cite{Cubizolles03,Hodby} are well explained by
employing a thermodynamic equilibrium model \cite{Petrov}. The
narrow $^6$Li resonance, traversed much more rapidly
\cite{Strecker03}, is not expected to fit such a description.  We
consider the zero temperature limit to model such experiments.  At low
temperatures one can use a single bosonic mode Hamiltonian
\cite{Javanainen04, Barankov04, Andreev04, Dukelsky04, Tikhonenkov05,
Miyakawa05,Mackie05} because of the Cooper instability which singles
out the zero momentum mode of the molecules produced.  Thus, we take
the Hamiltonian to be
\begin{eqnarray}
H&=&\sum_{\bfk,\sigma} \ek\cks^\dag\cks+{\cal E}(t)
\bz^\dag\bz\nonumber\\ 
~&~&+g\left(\sum_{\bf k} \cku\cmkd \bz^\dag + H.c.\right) \,,
\label{eq:ham}
\end{eqnarray}
where $\ek=\hbar^2k^2/2m$ is the kinetic energy of an atom with mass
$m$, and $g$ is the atom-molecule coupling strength.  The molecular
energy ${\cal E}(t) =\ef-\alpha t$ is linearly swept at a rate $\alpha >
0$ where $\ef$ denotes the fermi energy of the atoms, through resonance 
to induce adiabatic conversion of fermi atoms to
bose molecules.  The annihilation operators for the atoms, $\cks$,
obey fermionic anti-commutation relations, whereas the molecule
annihilation operator $\bz$ obeys a bosonic commutation relation.

\begin{figure}
\centering
\includegraphics[scale=0.5,angle=0]{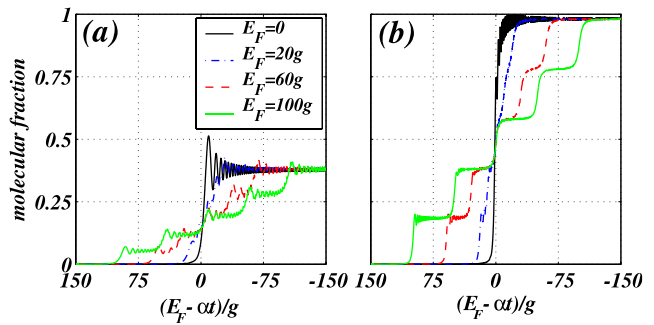}
\caption{(Color online) Many-body collective dynamics 
of adiabatic passage from a
fermionic atomic gas into a molecular Bose Einstein Condensate (BEC)
for five pairs of fermionic atoms.  (a) Sweep rate $\alpha = 2g^2
N$, (b) Sweep rate $\alpha = g^2 N/4$.  Overall efficiency is
independent of atomic dispersion in both (a) and (b).}
\label{fig2}
\end{figure}

The Hamiltonian can be further simplified by neglecting fermionic
dispersion.  This approximation has been commonly used
\cite{Mies00,Barankov04} and accounts for the use of a simple
two-level LZ model, as opposed to a multilevel one, for such systems.
To justify this assumption we have conducted many-body numerical
simulations to determine the effect of fermionic dispersion on the
adiabatic conversion efficiency.  Fig.~\ref{fig2} shows exact
numerical results for the adiabatic conversion of five atom pairs into
molecules, for different values of the atomic level spacing (and hence
of the Fermi energy $E_F$).  It demonstrates that the {\em final}
adiabatic conversion efficiency is completely insensitive to the
details of the atomic dispersion.  It is evident that, while
the exact dynamics depends on $E_F$, and the levels are sequentially
crossed as a function of time as the bound state crosses the level
energies, the same final efficiency is reached regardless of the
atomic motional timescale (i.e., regardless of level spacing).  In
particular, the figure shows that in the limit as $\alpha\rightarrow
0$, it is possible to convert {\em all} atom pairs into molecules.
This is a unique feature of the nonlinear parametric coupling between
atoms and molecules which should be contrasted with a marginal
conversion efficiency expected for linear coupling in the multi-level
LZ model.

Employing the degenerate model with $\ek=\epsilon$ for all $\bfk$
\cite{Tikhonenkov05, Miyakawa05, Mackie05}, it is convenient to
rewrite the Hamiltonian in terms of the following lowering, raising
and $z$-component operators \cite{Vardi01,Miyakawa05}:
$$
\jm=\frac{\bz^\dag\sum_{\bfk} \cku\cmkd}{(N/2)^{3/2}} ~,~
\jp=\frac{\sum_{\bfk} \cmkd^\dag\cku^\dag \bz}{(N/2)^{3/2}} \,,
$$
\begin{equation}
\jz=\frac{\sum_{\bfk,\sigma}\cks^\dag\cks-2\bz^\dag\bz}{N}\,,
\label{js}
\end{equation}
where $N=2\bz^\dag\bz+\sum_{\bfk,\sigma}\cks^\dag\cks$ is the
conserved total number of particles.  It is important to note that
$\jm,\jp,\jz$ do not span $SU(2)$, since the commutator $[\jp,\jm]$
yields a quadratic polynomial in $\jz$ (despite the fact that the
commutators $[\jp,\jz]$ and $[\jm,\jz]$ have the right commutation
relations).  The operators $\jx=\jp+\jm$ and
$\jy=-i(\jp-\jm)$ can also be defined.  Up to a $c$-number term,
Hamiltonian (\ref{eq:ham}) takes the form
\begin{equation}
H=\frac{N}{2}\left(\Delta(t)\jz+g{\frac{\sqrt{N}}{2}}\jx\right)\,,
\label{hamj}
\end{equation}
where $\Delta(t) = 2\epsilon-{\cal E}(t) = \alpha t$.  Defining a
rescaled time $\tau=\sqrt{N}gt$, and assuming a filled Fermi sea
(i.e., that the number of avialable fermionic states is equal to the
number of particles), we obtain the Heisenberg equations of motion,
\begin{eqnarray}
\frac{d}{d\tau}\jx &=& \delta(\tau)\jy \,, \nonumber \\
\frac{d}{d\tau}\jy &=& -\delta(\tau)\jx+\frac{3\sqrt{2}}{4}
\left(\jz-1\right)\left(\jz+\frac{1}{3}\right)\nonumber\\
~&~&-{\sqrt{2}\over{N}}\left(1+\jz\right)\,, \nonumber \\
\label{eom}
\frac{d}{d\tau}\jz&=&\sqrt{2}\jy\,,
\end{eqnarray}
which depend only on the scaled detuning $\delta(\tau) = \Delta(t)
/\sqrt{N}g = (\alpha/g^2N)\tau$.  It is interesting to note that
exactly these equations of motion are obtained for the two-mode
atom-molecule BEC \cite{Vardi01} where, for the bosonic case,
lowering, raising and $z$-component operators are defined as
$$
\jm=\frac{\bz^\dag \au \ad}{(N/2)^{3/2}} ~,~
\jp=\frac{ \ad^\dag\au^\dag \bz}{(N/2)^{3/2}}\,,
$$
\begin{equation}
\jz=\frac{2\bz^\dag\bz-\sum_{\bfk,\sigma}\aks^\dag\aks}{N}\,,
\label{eq:bosejs}
\end{equation}
where $\au$ and $\ad$ are bosonic annihilation operators obeying
bosonic commutation relations.  In these definitions, the sign of the
operator $\jz$ has been reversed relative to Eq.~(\ref{js}), mapping 
fermionic association to bosonic dissociation 
\cite{Tikhonenkov05,Miyakawa05,Mackie05}.

\section{Classical phase space}
\label{Sec:phase space}

The mean-field limit of Eqs.~(\ref{eom}) is given by replacing
$\jx,\jy$, and $\jz$ by their expectation values $u$, $v$, and $w$
which correspond to the real and imaginary parts of the atom-molecule
coherence and the atom-molecule population imbalance,
respectively.  Since quantum fluctuations in $\jz$ are of order $1/N$,
it is also consistent to omit the quantum noise term
$\sqrt{2}(1+\jz)/N$ in the equation for $d\jy/d\tau$ in (\ref{eom}), as
long as $\jz$ is of order unity.  For small $w$ however, when the
molecular population is of the order of its quantum fluctuations, this
quantum term becomes dominant and will have a significant effect on
sweep efficiency, as will be shown in
Sec.~\ref{sec:quantum_fluctuations}.
  
\begin{figure}
\centering
\includegraphics[scale=0.35]{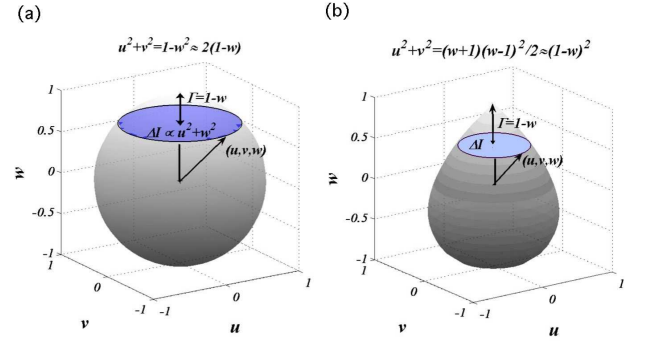}
\caption{(Color online) Two dimensional surfaces depicting 
classical phase space: 
(a) Bloch sphere of a two-mode Bose-Josephson system, as in 
Refs. \cite{Vardi01,Liu02} (b) zero single-particle entropy
surface of the atom-molecule system, as in Eqs.~(\ref{mfeom}).}
\label{fig4}
\end{figure}

In the classical field limit, the equations of motion
\begin{eqnarray}
\frac{d}{d\tau}u &=& \delta(\tau)v \,, \nonumber \\
\frac{d}{d\tau}v &=& -\delta(\tau)u+\frac{3\sqrt{2}}{4}
\left(w-1\right)\left(w+\frac{1}{3}\right) \,, \nonumber \\
\label{mfeom}
\frac{d}{d\tau}w&=&\sqrt{2}v \,,
\end{eqnarray}
depict the motion of a generalized Bloch vector on a two-dimensional
`tear-drop' shaped surface, (Fig.~\ref{fig4}b), determined by the
constraint,
\begin{equation}
u^2+v^2={1\over 2}(w-1)^2(w+1),
\label{eq:consrv}
\end{equation}
corresponding to the conservation of single-pair atom-molecule
coherence.  The peculiar shape of this equal-single-pair-entropy
surface is a result of the commutation relations for the operators
$\jm,\jp,\jz$.  For comparison, the two-level spin Hamiltonian may be
written only in terms of $SU(2)$ generators \cite{Vardi01,Liu02} and
the mean-field motion is restricted to the surface of a Bloch sphere
(see Fig.~\ref{fig4}a) as the mean-field approximation allows for the
factorization of the $SU(2)$ Casimir operator ${\bf J}^2 =
J_z(J_z-1)+J_+J_-=j(j+1)$ into the constraint
\begin{equation}
u^2+v^2+w^2=j^2 \,.
\end{equation}
The surface defined by the constraint (\ref{eq:consrv}) should be
viewed as the atom-molecule equivalent of this Bloch sphere.
Accordingly, we proceed by following the methods of Ref.~\cite{Liu02}
which correspond to a Bloch sphere like phase space, generalizing them
to the atom-molecule parametric coupling case.

Since the constraint of Eq.~(\ref{eq:consrv}) restricts the dynamics
to the two dimensional surface depicted in Fig. \ref{fig4}b, it is
readily seen that in the mean field limit, Hamiltonian (\ref{hamj}) 
is replaced by the classical form
\begin{equation}
H(w,\theta;\delta) = \frac{gN^{3/2}}{2}\left(\delta w +
\sqrt{(1+w)(1-w^2)} \cos\theta\right) \,.
\label{hamc}
\end{equation}
Here the Hamiltonian is expressed only in terms of the relative phase
between atoms and molecules $\theta \equiv \arctan(v/u)$, corresponding to
the azimuthal angle in Fig.~\ref{fig4}b, and the atom-molecule
population difference $w$, corresponding to its cylindrical axial
coordinate.

\begin{figure}
\centering
\includegraphics[scale=0.45]{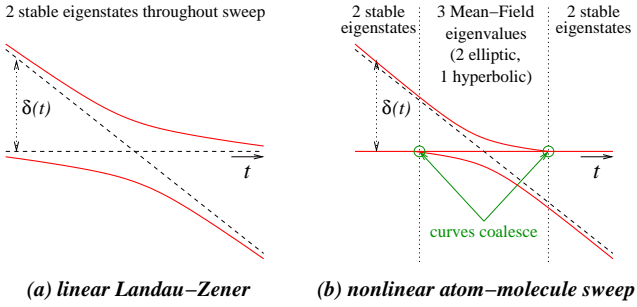}
\caption{(Color online) The adiabatic eigenvalues of a 
linear LZ problem (a) are
contrasted with the adiabatic level scheme of the atom-molecule 
nonlinear system (b). For a critical detuning, one of the  adiabatic 
eigenvalues in (b) splits into two, resulting in the emergence of an
additional hyperbolic fixed point.}
\label{fig5}
\end{figure}

The eigenvalues of the atom-molecule system at any given value of
$\delta$ correspond to the extrema $(w_0,\theta_0)$ of the classical 
Hamiltonian (\ref{hamc}), or equivalently, to the fixed points $(u_0,v_0,w_0)$
of the mean-field equations~(\ref{mfeom})]:
\begin{equation}
v_0 = 0~,~\frac{\sqrt{2}}{4}\left(w_0-1\right) \left(3w_0+1\right) =
\delta u_0 \,.
\label{fxp}
\end{equation}
The number of fixed points depends on the parameter $\delta$.  The
point $u_0=v_0=0,w_0=1$ is stationary for any value of $\delta$.
Using Eqs.~(\ref{eq:consrv}) and (\ref{fxp}), other fixed points
are found from the solutions of
\begin{equation}
\frac{(3w_0+1)^2}{4(w_0+1)}=\delta^2 \,,
\label{eq:roots}
\end{equation}
in the domain $w_0\in[-1,1]$.  Consequently, it is evident that for
$|\delta|\geq \sqrt{2}$ there are only two stationary solutions, one
of which is $w_0=1$.  However, this stationary point bifurcates at the
critical detuning of $\delta_c=\sqrt{2}$, so that for $|\delta| <
\delta_c$, there are three eigenstates, as depicted in
Fig.~\ref{fig5}b.  In contrast, in the linear LZ problem
(Fig.~\ref{fig5}a), eigenvalue crossings are avoided, and there are
only two eigenstates throughout.

\begin{figure}
\centering
\includegraphics[scale=0.45]{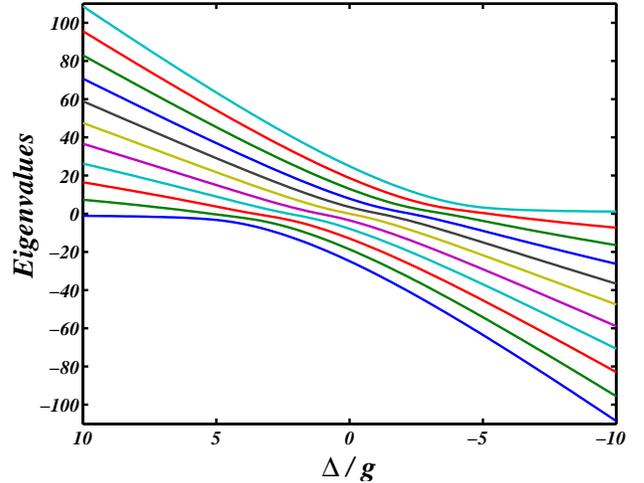}
\caption{(Color online) The ten lowest energy eigenvalues for Hamiltonian
(\ref{eq:ham}), drawn for the degenerate case, $\ek=0$.}
\label{fig6}
\end{figure}

The relation between the reduced sigle-particle picture of
Fig.~\ref{fig5}b and the full many-body sytem it approximately
represents, is illustrated in Fig.~\ref{fig6}, depicting numerically
calculated eigenvalues for ten atom pairs as a function of $\delta$,
when $E_F=0$.  One can envisage how, when adding more and more energy
levels, finally collapsing the levels to a single curve, the curve
structure of Fig.~\ref{fig5} emerges.  The bifurcation of the
all-atoms mode is shown to emerge from its quasi-degeneracy with
slightly higher many-body states with a few more molecules.

Stability analysis of the various fixed points, preformed by
linearization of the dynamical equations (\ref{eom}) about
$(u_0,v_0,w_0)$, yields the frequency $\Omega_0$ of small periodic
orbits around them:
\begin{equation}
\frac{\Omega_0}{g\sqrt{N}}=\sqrt{\delta^2+(1-3w_0)} \,.
\label{omegzz}
\end{equation}
From Eq.~(\ref{omegzz}) it is clear that the characteristic
oscillation frequency about the stationary point $(0,0,1)$ will be
$\sqrt{\delta^2-2}$.  Thus, for $\delta\geq\delta_c$ the point
$u_0=v_0=0,w_0=1$ is an elliptical fixed point, whereas for
$\delta\leq\delta_c$ it becomes a {\it hyperbolic} (unstable)
stationary point, with an immaginary perturbation frequency.  For the
remaining eigenvalues, we can use (\ref{eq:roots}) to obtain
\begin{equation}
\frac{\Omega_0}{g\sqrt{N}}=\sqrt{\frac{(1-w_0)(3w_0+5)}{4(w_0+1)}} \,,
\label{omegz}
\end{equation}
giving real $\Omega_0$ for all $w_0\in[-1,1]$, approaching 
zero as $w_o\rightarrow 1$.  Consequently, for
$|\delta|\leq\delta_c$ there are a total of two elliptical fixed
points, whereas for $|\delta|<\delta_c$ there are two elliptical and
one hyperbolic fixed points.

\begin{figure}
\centering
\includegraphics[scale=0.5,angle=-90]{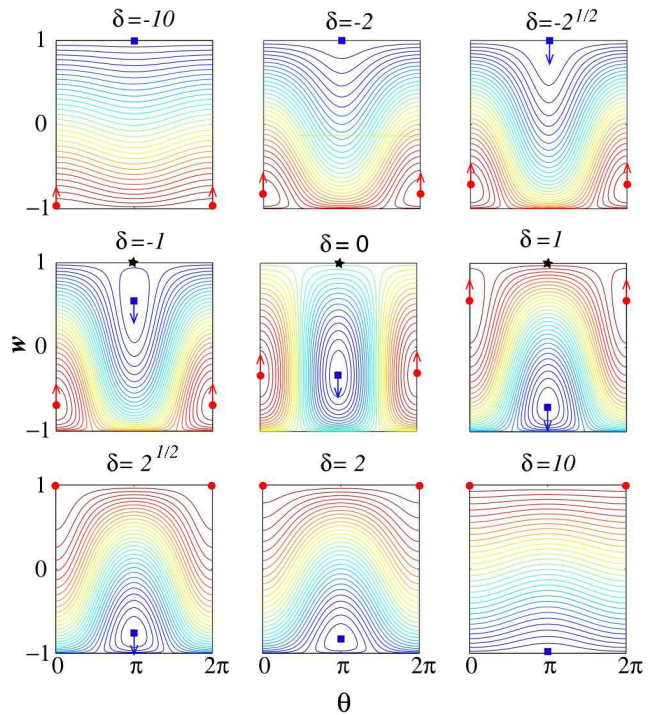}
\caption{(Color online) Equal-energy contours of Hamiltonian 
(\ref{hamc}) plotted as
a function of $w$ and $\theta$ for different detunings $\delta$.
$w=1$ is all atoms and $w=-1$ is all molecules.  The various fixed
points corresponding to adiabatic eigenvectors are marked by (blue)
squares, (red) circles and (black) stars.}
\label{fig7}
\end{figure}

The fixed point analysis carried out so far is summarized in
Fig.~\ref{fig7} where we plot the phase-space trajectories,
corresponding to equal-energy contours of Hamiltonian (\ref{hamc}),
for different values of $\delta$.  As expected from (\ref{hamc}), the
plots have the symmetry $(w,\theta;\delta) \leftrightarrow
(w,\theta+\pi;-\delta)$.  For sufficiently large detuning, $|\delta|>
\sqrt{2}$, Eq.~(\ref{eq:roots}) has only one solution in the range
$-1\leq w_0\leq 1$.  Therefore there are two (elliptic) fixed points,
denoted by a red circle corresponding to the solution of
Eq.~(\ref{eq:roots}), and a blue square at (0,0,1).  As the detuning
is changed, one of these fixed points (red circle) smoothly moves from
all-molecules towards the atomic mode.  At the critical detuning
$\delta=-\sqrt{2}$ a homoclinic orbit appears through the point
$(0,0,1)$ which bifurcates into an unstable (hyperbolic) fixed point
(black star) remaining on the atomic mode, and an elliptic fixed point
(blue square) which starts moving towards the molecular mode.
Consequently, in the regime $|\delta|<\sqrt{2}$ there are two elliptic
fixed points and one hyperbolic fixed point, corresponding to the
unstable all-atoms mode.  Another crossing occurs at $\delta=\sqrt{2}$
when the fixed point which started near the molecular mode (red
circle) coalesces with the all-atoms mode (black star).  Plotting the
energies of the fixed points as a function of detuning, one obtains
the adiabatic level scheme of Fig.~\ref{fig5}b.

As previously noted, for $|\delta|<\sqrt{2}$ the period of the homoclinic 
trajectory beginning at $(0,0,1)$ diverges. This divergence significantly
affects the efficiency of an adiabatic sweep through resonance because the 
linear response time to a perturbation in the Hamiltonian becomes infinitely
long. Consequently, the sweep is never truly adiabatic, and its expected 
efficiency is lower than the corresponding LZ efficiency. This effect 
is discussed in the following section. 

\section{Effect of fixed-point instability on adiabatic sweep
efficiency}
\label{sec:sweep}

Having characterized the classical phase-space structure for the
parametrically coupled atom-molecule system, we turn to the process
of adiabatically sweeping the detuning $\delta$ through resonance,
converting fermion atoms into bose molecules. As usual, adiabatic
following involves two typical timescales: the sweep time scale 
given by the inverse sweep rate $1/\alpha$ and the timescale associated 
with the period $1/\Omega_0$ of small periodic orbits around the fixed 
point, given in Eqs.~(\ref{omegzz}) and (\ref{omegz}). Slow
changes to the Hamiltonian (e.g., by variation of the detuning $\delta$)
shift the adiabatic eigenvalues, as depicted in Fig.~\ref{fig7}.
Starting with such an eigenvalue (e.g., the all-atoms mode for
an initial large negative $\delta$), the state of the system will only
be able to adiabatically follow the fixed point if its precession
frequency about it, $\Omega_0$, is large compared to the rate at which 
it moves. The adiabatic conversion efficiency is related to the
magnitude of this precession, which in turn is proportonal to the
classical action accumulated during the sweep.

The relation between the sweep conversion efficiency and the
accumulated classical action is illustrated in Fig.~\ref{fig4}.
Consider first the $SU(2)$ case of Fig.~\ref{fig4}a, where mean-field
motion is restricted to the surface of the Bloch sphere
$u^2+v^2+w^2=1$.  This illustration applies to the standard LZ case
\cite{Landau} as well as to the nonlinear Bose-Josephson system
\cite{Vardi01,Liu02,Garraway00}.  Having started from the `south pole'
$(0,0,-1)$ and carried out the sweep through resonance, the classical
state Bloch vector $(u,v,w)$ carries out small precessions about the
final adiabatic eigenvector which (for sufficiently large final
detuning) is parallel to the $w$ axis.  The surface-area enclosed
within this periodic trajectory is just the action, $\Delta I\propto
u^2+v^2$, accumulated during the sweep.  In the extreme limit of
perfect adiabatic following, the precession approaches a point
trajectory, having zero action.  Larger nonadiabaticity leads to
larger precession amplitude, and hence to larger accumulated action.
The remanent fraction in the initial state, $\Gamma=(1-w)$, is
directly related to $\Delta I$ by the conservation rule
$u^2+v^2=(1+w)(1-w)$, which near $w=1$ can be linearized to give
$\Delta I \approx 2 \Gamma$.  This is the expression usually used in
calculating LZ transition probabilities \cite{Landau,Liu02}.

For the atom-molecule parametric system of Fig.~\ref{fig4}b, the
situation is slightly different, since $u^2+v^2=(1+w)(1-w)^2$, leading
to a {\em square root} dependence of the remnent atomic fraction
$\Gamma$ on $\Delta I$ near $w=1$.  In order to estimate $\Delta I$,
we transform $w,\theta$ into action-angle variables $I,\phi$.  In
terms of these variables the non-adiabatic probability $\Gamma$ at any
finite sweep rate $\alpha$ is given by
\begin{equation}
\Gamma^2 = \frac{\Delta I}{2}={1\over 2}\int_{-\infty}^{\infty}
R(I,\phi) \, {\dot\Delta} \, \frac{d\phi}{{\dot\phi}} \,,
\label{eq:nonadiab}
\end{equation}
where $R(I,\phi)$ is related to the generating function of the
canonical transformation $(w,\theta) \rightarrow (I,\phi)$
\cite{Landau_L,Liu02}.

Equation (\ref{eq:nonadiab}) reflects our discussion on
characteristic timescales. In order to
attain adiabaticity, the rate of change of the adiabatic fixed points
through the variation of the adiabatic parameter $\Delta$,
$R(I,\phi)\, {\dot\Delta}$, should be slow with respect to the
characteristic precession frequency ${\dot\phi}=\Omega_0$ about these
stationary vectors. The action increament is proportional to
the ratio of these two timescales. 

\begin{figure}
\centering
\includegraphics[scale=0.6]{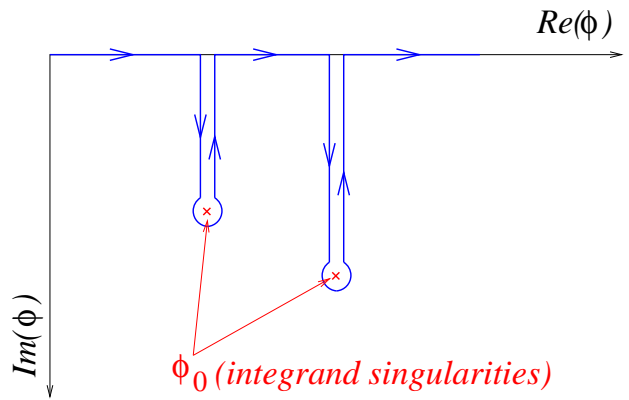}
\caption{(Color online) Contour of integration in LZ theory, 
for calculating the integral in Eq.~(\ref{eq:nonadiab}). 
All singularities lie off the real axis.}
\label{fig8}
\end{figure}

As long as ${\dot\phi}$ does not vanish, the accumulated action can be
minimized by decreasing ${\dot\Delta}$.  For a perfectly adiabatic
process where ${\dot\Delta}/{\dot\phi}\rightarrow 0$, the action is an
adiabatic invariant, so that a zero-action elliptic fixed point
evolves into a similar point trajectory.  For finite sweep rate, the
LZ prescription \cite{Landau} evaluates the integral in
(\ref{eq:nonadiab}) by integration in the complex plain, over the
contour of Fig.~\ref{fig8}, noting that the main contributions will
come from singular points, where ${\dot\phi}$ approaches zero and the
integrand diverges.  Since for a linear LZ system there are no
instabilities, all such singularities are guarenteed to lie off the
real axis, leading to the exponentially small LZ transition
probabilities.

The situation changes for nonlinear systems, where instabilities
arrise.  In section \ref{Sec:phase space} we have shown that for the
atom-molecule system with fermion atoms, the all-atoms mode becomes
unstable to association when the detuning hits the critical value of
$\delta_c=\sqrt{2}$.  From Eq.~(\ref{omegzz}) it is clear that the
characteristic frequency ${\dot\phi}=\Omega_0$ vanishes near $w_0=0$.
Consequently, there are singular points of the integrand in
(\ref{eq:nonadiab}) lying on the real axis.  In what follows, we show
that these poles on the real axis lead to power-law dependence of the
transfer efficiency on the sweep rate.
  
In order to evaluate the integral (\ref{eq:nonadiab}), we need to
investigate how the characteristic frequency ${\dot\phi}=\Omega_0$
depends on the action-angle $\phi$ near the instability point
$(u=0,v=0,w=1)$, where most action (and hence most nonadiabatic
correction) is accumulated.  It is evident from Eq.  (\ref{omegzz})
that the precession frequency near that point vanishes as
\begin{equation}   
\Omega_0\approx g\sqrt{N(1-w_0)} \,.
\label{omegzzz}
\end{equation}
Differentiating Eq.~(\ref{eq:roots}) with respect to $t$, we find that
the rate of change of adiabatic fixed-points due to a linear sweep is,
\begin{equation}
{{\dot w}_0}=\frac{4\alpha}{g\sqrt{N}}\frac{(w_0+1)^{3/2}}{3w_0+5} \,.
\label{wdot}
\end{equation}
Having found ${{\dot w}_0}$, we can now find the explicit form
of the transformation from $w_0$ to the action-angle variable
$\phi$, near the instability. The action-angle may be written as
\begin{equation}
\phi=\int {\dot\phi} dt =\int \Omega_0\frac{dw_0}{{\dot w}_0} \,.
\label{phitow}
\end{equation}  
In the vicinity of the singularity ($w_0=1$) we have
$\Omega_0 \approx g\sqrt{N(1-w_0)}$ and ${\dot w}_0\approx
{\sqrt{2}\alpha/g\sqrt{N}}$, resulting in
\begin{equation}
\phi=\frac{g^2 N}{\alpha}\frac{\sqrt{2}}{3}(1-w_0)^{3/2} \,.
\label{phiofw}
\end{equation}
Thus, as the adiabatic eigenstate approaches $w_0=1$ the angle
$\phi$ vanishes as $(1-w_0)^{3/2}$ whereas the characteristic
frequency ${\dot\phi}$ approaches zero as $(1-w_0)^{1/2}$.
Consequently, we finally find from (\ref{omegzzz}) and (\ref{phiofw})
that near the singularity, $\dot\phi$ is given in terms of $\phi$ as
\begin{equation}
{\dot\phi}=\left(3\sqrt{N\over 2} g\alpha\right)^{1/3}\phi^{1/3} \,.
\label{dotphiofphi}
\end{equation}
Substituting (\ref{dotphiofphi}) and $\dot\Delta=\alpha$ into Eq.
(\ref{eq:nonadiab}) we find that the nonadiabatic correction  depends on
$\alpha$ as
\begin{equation}
\Gamma \propto \alpha^{1/3} \,.
\label{cubicroot}
\end{equation}

\section{Quantum Fluctuations} \label{sec:quantum_fluctuations}

So far, we have neglected the effect of quantum fluctuations, which
may be partially accounted for by the $c$-number limit of the source
term $(\sqrt{2}/N)(1+\jz)$ in Eqs.~(\ref{eom}).  As a result, we found
in the previous section that $\dot w_0$ does not vanish as $w_0$
approaches unity, and the remaining atomic population scales as the
cubic root of the sweep rate if the initial average molecular fraction
is larger than the quantum noise.  However, starting purely with
fermion atoms (or with molecules made of bosonic atoms), corresponding
to an unstable fixed point of the classical phase space, fluctuations
will serve to trigger the association process and will thus initially
dominate the conversion dynamics.

\begin{figure}
\centering
\includegraphics[scale=0.45]{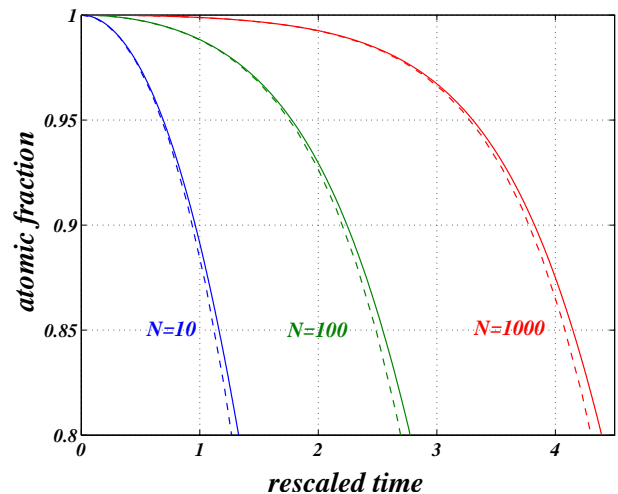}
\caption{(Color online) Atomic population fraction versus time, 
starting with a gas of fermion atoms using exact $N=10,100,1000$ 
particle calculations (solid lines) and the mean-field theory of 
Eqs.~(\ref{mfeomnoise}) (dashed lines), with $\delta=0$.}
\label{fig9}
\end{figure}

In order to verify that such quantum fluctuations can be accurately
reproduced by a `classical' noise term near the unstable fixed point
$w=1$, we compare the onset of instability from exact many-body
calculations to the onset of mean-field instability according to the
revised mean-field equations,
\begin{eqnarray}
\frac{d}{d\tau}u &=& \delta(\tau)v \,, \nonumber \\
\frac{d}{d\tau}v &=& -\delta(\tau)u+\frac{3\sqrt{2}}{4}
\left(w-1\right)\left(w+\frac{1}{3}\right) \nonumber \\
~&~&+\frac{\sqrt{2}}{N}(1+w)  \,, \nonumber \\
\label{mfeomnoise}
\frac{d}{d\tau}w&=&\sqrt{2}v \,,
\end{eqnarray}
where we have retained the ${\cal O}(1/N)$ noise term
$(\sqrt{2}/N)(1+w)$.  The results, shown in Fig.~\ref{fig9}, show
excellent agreement in the early-time dynamics, indicating that the
mean-field noise term gives the correct behavior near the instability
point.

Having established the accuracy of the noise term in
Eqs.~(\ref{mfeomnoise}) we proceed to investigate its effect on sweep
efficiencies.  When this additional term is accounted for,
Eq.~(\ref{eq:roots}) must be replaced by
\begin{equation}
\delta=\frac{2}{\sqrt{w_0+1}}\left(\frac{3w_0+1}{4}-\frac{w_0+1}{N(w_0-1)} 
\right) \,.
\label{eq:roots_noise}
\end{equation}
This expression reduces to our previous result in Eq.~(\ref{eq:roots})
when the second term on the r.h.s. of Eq.~(\ref{eq:roots_noise}),
resulting from the quantum fluctuations, can be neglected compared to
the first term.  Since $\frac{3w_0+1}{4}$ is of order unity around
$w_0=1$, our previous treatment is only valid provided that
$|w_0(t_i)-1|\gg 1/N$.

For smaller initial molecular population, Eq.~(\ref{wdot}) should be
replaced by
\begin{equation}
{{\dot w}_0}=\frac{\alpha}{g\sqrt{N}}\left/\left[
\frac{3w_0+5}{4(w_0+1)^{3/2}}+\frac{w_0+3}{N(w_0+1)^{1/2}(w_0-1)^2}
\right]\right. .
\end{equation}
Hence, in the vicinity of $w_0=1$ the eigenvector velocity in the $w$
direction vanishes as 
\begin{equation}
{{\dot w}_0} = (\sqrt{N}\alpha/g\sqrt{8}) \left(w_0-1\right)^2 \,.
\label{wdot_noise}
\end{equation}  
In contrast to the nonvanishing eigenvalue velocity in
Eq.~(\ref{wdot}).  Substituting ${\dot w}_0$ from
Eq.~(\ref{wdot_noise}) into Eq.~(\ref{phitow}), we obtain the
action-angle,
\begin{equation}
\phi=-\frac{\sqrt{32}g^2}{\alpha}(w_0-1)^{-1/2} \,.
\end{equation} 
The characteristic frequency $\dot\phi$ is now proportional to
$(\alpha\phi)^{-1}$ instead of Eq.~(\ref{dotphiofphi}) so that $\Delta
I\propto\alpha^2$, and \cite{Ishkhanyan04,Altman05,Barankov05}
\begin{equation}
\Gamma \propto \alpha~.
\label{linear}
\end{equation}

Equations (\ref{linear}) and (\ref{cubicroot}) constitute the main
results of this work.  We predict that the remnant atomic fraction in
adiabatic Feshbach sweep experiments scales as a power-law with
sweep rate due to the curve crossing in the nonlinear case.  When the
system is allowed to go near the critical point (i.e., when
$1-w_0(t_i)\ll 1/N$) quantum fluctuations are the major source of
non-adiabtic corrections, leading to a linear dependence of the
remnent atomic fraction on the sweep rate.  We note that a similar
linear dependence was predicted for adiabatic passage from bosonic
atoms into a molecular BEC \cite{Ishkhanyan04}.  When the initial
state is such that it has already a large molecular population (i.e.
for $1-w_0(t_i) \gg 1/N$) and fluctuations can be neglected, we obtain
a cubic-root dependence of the the final atomic fraction on sweep
rate.

\begin{figure}
\centering
\includegraphics[scale=0.5,angle=-90]{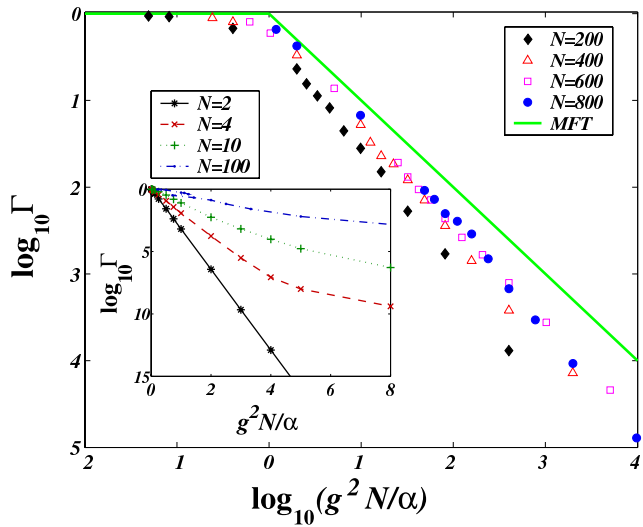}
\caption{(Color online) Many-body calculations for the fraction of 
remnant atoms, $\Gamma$, versus dimensionless inverse sweep rate 
for various particle numbers in the range $N = 2$ to 800. 
The many-body results for large number of particles converge to 
the mean-field results (solid green line) computed numerically from 
the mean-field limit of Eqs.~(\ref{eom}).}
\label{fig3}
\end{figure}

\section{Numerical Many-Body Results, and Comparison to Experiment}
\label{sec:numerical}

In order to confirm the predictions of Section~\ref{sec:sweep} and
Section~\ref{sec:quantum_fluctuations}, we carried out excat many-body
numerical calculation for particle numbers in the range {$2\le N \le
800$}, by Fock-space representation of the operators ${\cal J}_i$ and
direct propagation of the many-body equations (\ref{eom}), according
to the methodology of \cite{Tikhonenkov05}.  Fig.~\ref{fig3} shows
$\Gamma$ versus dimensionless inverse sweep rate ${g^2 N/\alpha}$.
The exact calculations are compared with a mean-field curve (solid
green line), computed numerically from the mean-field
equations~(\ref{mfeomnoise}).  The log-log plot highlights the mean-field
power-law dependence, obtained in the slow ramp regime $\alpha<g^2 N$,
whereas the log-linear insert plot demonstrates exponential behavior.
For a single pair of particles, $N=2$, the quantum association problem
is formally identical to the linear LZ paradigm, leading to an
exponential dependence of $\Gamma$ on sweep rate (see insert of
Fig.~\ref{fig1}).  However, as the number of particles increases,
many-body effects come into play, and there is a smooth transition to
a power-law behavior in the slow ramp regime $\alpha<g^2 N$.  We note
that this is precisely the regime where Eq.~(\ref{eq:nonadiab}) can be
used to estimate $\Delta I$ and $\Gamma$ \cite{Landau}.  The many-body
calculations converge to the mean-field limit, corresponding to a
linear dependence of $\Gamma$ on $\alpha$, as predicted in Eq.
(\ref{linear}).

The results shown in Fig.~\ref{fig3} prove the convergence of
many-body calculations to the mean-field theory used as a basis to our
analysis in previous sections.  Having established the validity of
this classical field theory, and numerically confirmed the appearance
of power-law behavior, we return to the experimental results of
Ref.~\cite{Strecker03} shown in Fig.~\ref{fig1}.  Comparison of our
mean-field numerical calculation with the experimental data (red
squares in Fig.~\ref{fig1}) clearly shows good agreement.  However,
since an equally good exponential fit can be found \cite{Strecker03},
as shown in the insert of Fig.~\ref{fig1} (dashed line), current
experimental data does not serve to determine which of the alternative
theories is more appropriate.  We have obtained similar agreement with
the experimental data of Ref.~\cite{Regal03}, but data scatter and
error bars are again too large to conclusively resolve power-laws from
exponentials.  Additional precise experimental data over a wider range
of slow ramp sweep rates and different particle numbers will be
required to verify or to refute our theory.

\section{Summary and Discussion}
\label{sec:summary}

We have shown that nonlinear effects can play a significant role in
the atom-molecule conversion process for degenerate fermionic atomic
gases.  In linear LZ theory, the precession of the two-state Bloch
vector about the adiabatic eigenstate never stops (all the poles of
the the integrand in Eq.~(\ref{eq:nonadiab}) lie off the real axis),
leading to exponential dependence on sweep rate.  The nonlinear nature
of the reduced mean-field dynamics in the large $N$ limit of the
many-body system, introduces dynamical instabilities.  For the
fermionic association case, it is the all-atoms mode that becomes
unstable.  The period of the precession about this mode diverges,
leading to {\it real} singularities of the integrand in
(\ref{eq:nonadiab}).  Consequently, power-law nonadiabaticity is
obtained.

While the experimental data was originally fit with LZ exponential
behavior \cite{Strecker03}, it fits a power-law dependence just as
well.  Future experimental work with a larger range of sweep rates,
should serve to determine which fit is best at low temperatures.

The modification of LZ behavior into a $\Gamma\propto\alpha^{3/4}$ 
power-law dependence, has been predicted for linearly-coupled, 
interacting Bose-Josephson systems \cite{Liu02,Garraway00}.  
Here, and in Ref.~\cite{Pazy05}, we applied the theoretical 
technique of \cite{Liu02}, adapting it to the case of a non-spherical 
two dimensional phase space surface. The exact power-law
for the atom-molecule system was shown to depend on the role
quantum noise plays in the conversion process. When it is negligible,
we find that $\Gamma\propto\alpha^{1/3}$. However, starting from 
a purely atomic gas, quantum fluctuations dominate the dynamics,
resulting in a $\Gamma\propto\alpha$ power law. The same linear 
dependence was previously found for  the bosonic photoassociation 
problem \cite{Ishkhanyan04}. 

Our numerical results support our analytical predictions,
demonstrating how exponential LZ behavior, applicable to two atoms, is
transformed into a power law dependence as the number of atoms
increases (see Fig.~\ref{fig3}).  The analysis based on
Eq.~(\ref{eq:nonadiab}) makes the differences between the two cases
transparent, relating them to different types of singularities.

We note that the same power laws of $\Gamma\propto\alpha^{1/3}$ and
$\Gamma\propto\alpha$ appear in recent theoretical studies of
dynamical projection onto Feshbach molecules
\cite{Altman05,Barankov05}.  The power-law in \cite{Altman05} results
from the nature of the projected pairs, correlated pairs giving a
$\Gamma\propto\alpha^{1/3}$ power law whereas uncorrelated pairs give
linear dependence, due to their respective overlaps with the molecular
state.  For comparison, the quantum noise term in our analysis
corresponds to initial uncorrelated spontaneous emission, leading to
$\Gamma\propto\alpha$ linear behavior, whereas for a larger initial
molecular population, correlations between emitted pairs are
established and emission becomes coherent, yielding the
$\Gamma\propto\alpha^{1/3}$ power law.  The approach taken in
\cite{Barankov05} is rather different, based on a variant of the
Wiener-Hopf method, yet it results in precisely the same power-laws.

\begin{acknowledgments}
We gratefully acknowledge support for this work
from the Minerva Foundation through a grant for a Minerva Junior 
Research Group,the U.S.-Israel Binational Science Foundation 
(grant No 2002147), the Israel Science Foundation for a Center 
of Excellence (grant No.~8006/03), and the German Federal 
Ministry of Education and Research (BMBF) through the DIP project.
\end{acknowledgments}

\pagebreak


\end{document}